\documentclass{ws-ijmpcs}

\newcommand{\be}{\begin{equation}}
\newcommand{\ee}{\end{equation}}
\newcommand{\bea}{\begin{eqnarray}}
\newcommand{\eea}{\end{eqnarray}}
\newcommand{\nn}{\nonumber}

\begin{document}

\markboth{Abalo, Milton, and Kaplan}
{Scalar Casimir Energies of Tetrahedra }

\catchline{}{}{}{}{}
%
%%%%%%%%%%%%%%%%%%%%%%%%%%%%%%%%%%%%%%%%%%%%%%%%%%%%%%%%%%%%%%%%%%%%

\title{SCALAR CASIMIR ENERGIES OF TETRAHEDRA}

\author{E. K. ABALO, K. A. MILTON }

\address{Homer L. Dodge Department of Physics and Astronomy, University of Oklahoma\\ Norman, OK 73019, USA
\\
abalo@nhn.ou.edu, milton@nhn.ou.edu}

\author{L. KAPLAN}

\address{Department of Physics, Tulane University\\
New Orleans, LA 70118, USA\\
lkaplan@tulane.edu}

\maketitle

\begin{history}
\received{Day Month Year}
\revised{Day Month Year}
\end{history}

\begin{abstract}
New results for scalar Casimir self-energies arising from interior modes are presented for the 
three integrable tetrahedral cavities. Since the eigenmodes are all known, the energies
can be directly evaluated by mode summation, with a point-splitting regulator, which amounts
to evaluation of the cylinder kernel.  The correct Weyl divergences, depending on the volume,
surface area, and the corners, are obtained, which is strong evidence that the counting
of modes is correct.  Because there is no curvature, the finite part of the quantum energy
may be unambiguously extracted. Dirichlet and Neumann boundary conditions are considered and systematic behavior of the energy in terms of geometric invariants
is explored.
\keywords{Casimir energy; Casimir self-energy}
\end{abstract}

\ccode{PACS numbers: 13.70.+k,11.10.Gh,42.50.Lc,42.50.Pq}

\section{Introduction}	

The concept of a self-energy in the context of the Casimir effect remains elusive. 
Since 1948, the year of H. B. G. Casimir's seminal paper,\cite{casimir} what is now 
called the Casimir effect has captivated many. Yet, while Casimir later predicted an 
attractive force for a spherical conducting shell,\cite{casimir53}
Boyer proved the self-stress in that case to be instead repulsive,
 which was an even more unexpected result.\cite{boyer} 
Since Boyer's formidable calculation, many other configurations were examined: 
cylinders, boxes, wedges, etc. The literature abounds with these results;
for a review see Ref.~\refcite{milton-rev}.  However, since there are other
well-known cases of cavities where the interior modes are known exactly,
it is surprising that essentially no attention had been paid to these.
For example, recently we presented the first results for Casimir self-energies
for cylinders of equilateral, hemiequilateral, and right isosceles triangular
cross sections,\cite{Cylpaper} even though the spectrum is well-known and appears in general
textbooks.\cite{radbook,embook}  
Possibly, the reason for this neglect was that only interior modes could 
be included for any of these cases, unlike the case of
a circular cylinder, where both interior and exterior modes must be included in
order to obtain a finite self-energy.  However, the extensive attention to
rectangular cavities puts the lie to this hypothesis.\cite{lukosz}\cdash\cite{ambjorn}
It seems not to have been generally appreciated that finite results can be obtained
in all these cases because there are no curvature divergences for boxes constructed
from plane surfaces.

In Ref.~\refcite{Cylpaper} we obtained exact, closed-form results for the three-mentioned
integrable triangles, both in a plane, and for cylinders with the corresponding cross section.
The expected Weyl divergences related to the volume, surface area, and the corners of the
triangle were obtained, going a long way toward verifying the counting of modes, which
is the most difficult aspect of these calculations.  Moreover, we were able to successfully
interpolate between the results for these triangles by using an efficient numerical evaluation,
and showed that the results, for Dirichlet, Neumann, and perfect conducting boundary conditions,
lie on a smooth curve, which was reasonably well-approximated by the result of a proximity
force calculation.  In this paper we show that the same techniques can be applied to 
tetrahedral boxes; again, there are exactly three integrable cases, where an explicit
spectrum can be written down. Again, it is surprising that the Casimir energies for
these cases are not well-known.  The only treatment of a pyramidal box found in the Casimir energy 
literature appears in a relatively unknown work of Ahmedov and Duru,\cite{AhmedovSmT} which
however, seems to contain a counting error.

In this paper we present Casimir energy calculations for three integrable tetrahedra.
For each cavity we consider a massless scalar field subject to Dirichlet and Neumann boundary conditions
on the surfaces. We regulate the mode summation by temporal point-splitting, which amounts to 
evaluation of what is called the cylinder kernel,\cite{Fulling:2003zx} and extract both divergent
(as the regulator goes to zero) and the finite contributions to the energy. In the end, we explore the functional behavior 
of the Casimir energies with respect to an appropriately chosen ratio of the cavities' 
volumes and surface areas, including cubic and spherical 
geometries with the same boundary conditions. Further details will appear in Ref.~\refcite{Tetra}.

\subsection{Point-splitting regularization}
We regularize our results by temporal point-splitting. As explained in 
Ref.~\refcite{Cylpaper}, after a Euclidean rotation, we obtain
\begin{equation}
\mathcal{E}=\frac{1}{2}\lim_{\tau \rightarrow 0}\left(-\frac{d}{d\tau}\right)\sum_{k,m,n}
 e^{-\tau \sqrt{\lambda_{kmn}^2}}\,\,,
\end{equation}
where the sum is over the quantum numbers that characterize the eigenvalues, and $\tau$ is
the Euclidean time-splitting parameter, supposed to tend to zero at the end of the calculation.
One recognizes the sum as the integrated cylinder kernel.\cite{Fulling:2003zx}
Next, we proceed to re-express the sum with Poisson's summation formula.

\subsection{Poisson resummation}
Poisson's summation formula allows one to recast a slowly convergent sum into a more rapidly 
convergent sum of its Fourier transform,
\begin{equation}
 \sum_{m=-\infty}^\infty f(m)=\sum_{n=-\infty}^\infty\left(\int_{-\infty}^{\infty}e^{2 \pi i m n}f(m)\,dm\right)\,.
\label{psf}
\end{equation}
By point-splitting and resumming, we are able to isolate the finite parts, which are Casimir self-energies,
and the corresponding divergent parts, which are the the Weyl terms.
The resummed expressions below are obtained by evaluating the Fourier transforms in spherical coordinates.\cite{Cylpaper,lukosz} They are used in regularizing expressions throughout the paper.
\begin{align}
\bigg(-\frac{d}{d\tau}\bigg)\sum_{p,q,r}
e^{-\tau\sqrt{\alpha(p+a)^2+\beta(q+b)^2+\gamma(r+c)^2}}
= &\frac{24\pi}{\sqrt{\alpha\beta\gamma}\,\tau^4}
-\frac{1}{2\pi^3\sqrt{\alpha\beta\gamma}}\,
\\&\times\sideset{}{'}
\sum_{p,q,r}\bigg(
\frac{e^{-2\pi i(pa+qb+rc)}}{(p^2/\alpha + q^2/\beta + r^2/\gamma)^2}\bigg),\nn
\end{align}
\be
\bigg(-\frac{d}{d\tau}\bigg)\sum_{p,q}
e^{-\tau\sqrt{\alpha(p+a)^2+\beta(q+b)^2}}
=\frac{4\pi}{\sqrt{\alpha\beta}\,\tau^3}
-\frac{1}{4\pi^2\sqrt{\alpha\beta}}\,\sideset{}{'}\sum_{p,q}
\frac{e^{-2\pi i(pa+qb)}}{(p^2/\alpha + q^2/\beta)^{3/2}},
\ee
\be
\bigg(-\frac{d}{d\tau}\bigg)\sum_{p}
e^{-\tau\sqrt{\alpha(p+a)^2}}=\frac{2}{\sqrt{\alpha}\,\tau^2}
-\frac{\sqrt{\alpha}}{2\pi^2}\,\sideset{}{'}\sum_{p}\frac{e^{-2\pi i (pa)}}
{p^2}.
\ee
Here the prime means that all positive and negative integers are included
in the sum, but not the case when all are zero.

\section{Casimir Energies of Tetrahedra}
The three integrable tetrahedra mentioned above are not recent discoveries. 
They have, in fact, been the subject of a few articles.\cite{TerrasSwanson}\cdash\cite{Krishnamurthy} 
However, there appears to be only one Casimir energy article concerning one of these tetrahedra, 
which we denote as the ``small" tetrahedron.\cite{AhmedovSmT} These tetrahedra are integrable
in the sense that their eigenvalue spectra are known explicitly, and there are no other such
tetrahedra. We will successively look at the ``large," ``medium," and ``small" tetrahedra, as
defined below, and obtain interior scalar Casimir energies for Dirichlet and Neumann boundary conditions.
Although the exterior problems cannot be solved in these cases, 
the finite part of the  interior energies are well defined
because the curvature is zero, and hence the second heat kernel coefficient vanishes.

\subsection{Large tetrahedron}%------------------------------------------------------------
\begin{figure}[h]	
	\centering
\includegraphics[scale=0.7, trim= 5cm 10cm 5cm 10cm, clip]{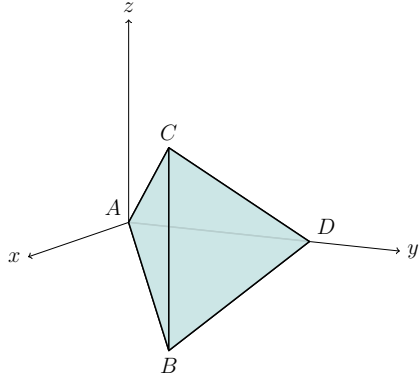}
	\caption{Large tetrahedron: $-x<z<x$ and $x<y<2a-x$. }
	\label{LargeTfigure}
	\end{figure}
The first tetrahedron, sketched in Fig.~\ref{LargeTfigure},
 which we denote ``large," is comparatively the largest or rather the most symmetrical. 
One can obtain a medium tetrahedron by bisecting a large tetradron and idem for the 
small and medium tetrahedra. One should note that the terms ``large,"
 ``medium," and ``small''
are merely labels, since one can always rescale each tetrahedron independently 
of the others. The spectrum and complete eigenfunction set for the large
tetrahedron, as well as those of the other tetrahedra, are known and appear in Ref.~\refcite{TerrasSwanson},
\be \label{LTeigval}
\lambda_{kmn}^2=\frac{\pi^2}{4a^2}\left[3(k^2+m^2+n^2)-2(km+kn+mn)\right]\,.
\ee
With Dirichlet and Neumann boundary conditions, different constraints are imposed on the spectrum,
that is, on the ranges of the integers $k$, $m$, and $n$.
		
\subsubsection{Dirichlet BC}
The complete set of modes for Dirichlet boundary conditions is given by the restrictions $0<k<m<n$. 
After extending the sums to all of $(k,m,n)$-space and removing unphysical terms, the 
Dirichlet Casimir energy for the large tetrahedron can be defined in terms of the function
\be \label{gfunction}
g(p,q,r)=e^{-\tau
\sqrt{(\pi/a)^2\left(p^2+q^2+r^2\right)}}\,,
\ee
and written as
\begin{align}
E=\frac{1}{48}\lim_{\tau \rightarrow 0}\left(-\frac{d}{d\tau}\right)\sum_{p,q,r}
\Big[ & g(p,q,r)+g(p+1/2,q+1/2,r+1/2)-6\,g(p,q,q)
\\& -6\,g(p+1/2,q+1/2,q+1/2)+8\,g(\sqrt{3}p/2,0,0) \nn
\\& +3\,g(p,0,0) \Big]\nn,
\end{align}
where the sums extend over all positive and negative integers including zero.
(In the third and fourth terms only $p$ and $q$ are summed over,
while in the last two terms only $p$ is summed.)
Note that the time-splitting has automatically regularized the sums, and it is easy
to extract the finite part,
\begin{align}
E_{L}^{(D)}=\frac{1}{a}\bigg( & -\frac{1}{96 
\pi^2}\big[Z_3(2;1,1,1)+Z_{3b}(2;1,1,1)\big]+\frac{1}{8\pi}
\zeta(3/2)L_{-8}(3/2)\
\\&+\frac{1}{16\pi} Z_{2b}(3/2;2,1)-\frac{\pi}{96}-\frac{\pi\sqrt{3}}{72}\, \bigg), \nn
\end{align}
where\cite{GlasserZucker}
 \be
 \sideset{}{'}\sum_{m,n}(m^2+2 n^2)^{-s}=2\zeta(s)L_{-8}(s)\,.
 \ee
The energy then evaluates numerically to
 \be
E_{L}^{(D)} =-\frac{0.0468804266}{a}\,.
\ee	
The function $L_{-8}$ is a  Dirichlet $L$-series, which are defined as $L_k(s)= \sum_{n=1}^{\infty}\chi_k(n)\,n^{-s}$ where $\chi_{k}$ 
is the number-theoretic character.\cite{GlasserZucker} The Epstein zeta functions $Z_3$, $Z_{3b}$, and $Z_{2b}$ are defined respectively as  
\be
	Z_3(s;a,b,c)=\sideset{}{'}\sum_{k,m,n}(a\,k^2+b\,m^2+c\,n^2)^{-s},
\ee
\be
Z_{3b}(s;a,b,c)=\sideset{}{'}
\sum_{k,m,n}(-1)^{k+m+n}(a\,k^2+b\,m^2+c\,n^2)^{-s},
\ee
and
\be
	Z_{2b}(s;a,b)=\sideset{}{'}\sum_{m,n}(-1)^{m+n}(a\,m^2+b\,n^2)^{-s}.
\ee

 The divergent parts, also extracted from the regularization procedure, 
follow the expected form of Weyl's law with the quartic divergence associated with the volume $V$, 
the cubic divergence associated with the surface area $S$, and the quadratic 
divergence matched with the corner coefficient
\be
E^{(D)}_{\rm div}= \frac{3 V}{2 \pi^2 
\tau^4}-\frac{S}{8\pi\tau^3}+\frac{C}			
{48\pi\tau^2}. 
\ee
Here and subsequently, the corner coefficient $C$ for a polyhedron is defined as\cite{Fedosov}
\be
C = \sum_j\left(\frac{\pi}{\alpha_j}-\frac{\alpha_j}{\pi}\right)L_j\,,
\ee
where the $\alpha_j$ are dihedral angles and the $L_j$ are the corresponding edge lengths. 
The above expression for the divergences will be the same for all subsequent cavities with 
Dirichlet boundary conditions.

\subsubsection{Neumann BC}
In the case of Neumann boundary conditions, the complete set of mode numbers must satisfy
 $0 \leq k \leq m \leq n$, excluding the case when all mode numbers are zero. 
The Neumann Casimir energy can be defined in terms of the preceding Dirichlet result as
\be
E_L^{(N)}=E_L^{(D)}-\frac{1}{8\pi a}\Big[2\,\zeta(3/2)L_{-8}(3/2)+Z_{2b}(3/2;2,1) \Big],
\ee
which gives us a numerical value of
\be
E_{L}^{(N)} =-\frac{0.1964621484}{a}\,.
\ee	
The divergent parts also match the expected Weyl terms for Neumann boundary conditions. 
We note that the cubic divergence's coefficient changes sign when comparing Dirichlet and 
Neumann divergent parts:
\be
E^{(N)}_{\rm div}=
 \frac{3 V}{2 \pi^2 \tau^4}
+\frac{S}{8\pi\tau^3}+\frac{C}{48\pi\tau^2}.
\ee
This form is also obtained for all the following calculations involving Neumann boundary conditions.

\subsection{Medium tetrahedron}%------------------------------------------------------
\begin{figure}[h]
	\centering
	\includegraphics[scale=0.7, trim= 5cm 11cm 5cm 11cm, clip]{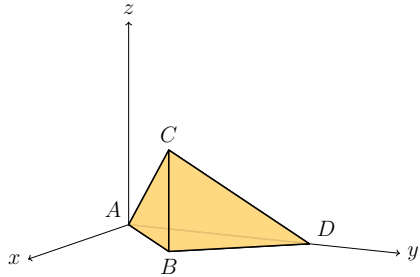}
	\caption{\label{medtetfig}Medium tetrahedron: $0<z<x$ and $x<y<2a-x$. }
\end{figure}
The eigenvalue spectrum of the medium tetrahedron,
shown in Fig.~\ref{medtetfig}, is of the same form as that of the large tetrahedron 
[Eq.~\eqref{LTeigval}] with different constraints.
\subsubsection{Dirichlet BC}
The complete set of mode numbers for the Dirichlet case satisfies the constraints $0<m<n<k<m+n$. Following the same regularization procedure used in the preceding cases, we obtain the 
Dirichlet Casimir energy in terms of the Dirichlet result for the large tetrahedron,
\begin{align}
E_{M}^{(D)} & = \frac{1}{2}\, E_{L}^{(D)}+\frac{1}{96\pi 
a}\Big[3\,\zeta(3/2)\beta(3/2)-(1+\sqrt{2})\pi^2\Big] ,
\end{align}	
where we used\cite{GlasserZucker}	
\be
 \sideset{}{'}\sum_{m,n}(m^2+n^2)^{-s}=4\zeta(s)\beta(s)\,.
\ee
The Casimir energy evaluates to
\be
E_{M}^{(D)} =-\frac{0.0799803933}{a}\,.
\ee	
Here the function $\beta$ is the Dirichlet beta function, also known as $L_{-4}$, defined as $\beta(s)=\sum_{n=0}^{\infty}(-1)^{n}(2n+1)^{-s}$.

\subsubsection{Neumann BC}
With Neumann boundary conditions, the complete set of mode numbers is restricted to 
$0 \leq m \leq n \leq k \leq m+n$, excluding the all-null case. 
As with the Dirichlet case, the Neumann Casimir energy for the medium 
tetrahedron can be expressed in terms of the Neumann result for the large tetrahedron:
\begin{align}
E_{M}^{(N)} & = \frac{1}{2}\, E_{L}^{(N)}-\frac{1}{96\pi 
a}\Big[3\,\zeta(3/2)\beta(3/2)+(1+\sqrt{2})\pi^2\Big] \nn
\\& = -\frac{0.1997008024}{a}\,.
\end{align}

\subsection{Small tetrahedron}%------------------------------------------------
\begin{figure}[h]
	\centering
	\includegraphics[scale=0.7, trim= 5cm 11cm 5cm 11cm, clip]{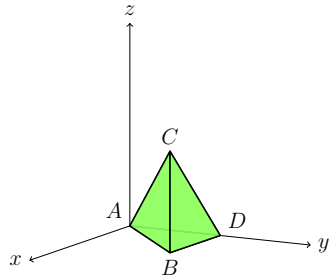}
	\caption{\label{smtetfig}Small tetrahedron:  $0<z<x$ and $0<y<a$.}
\end{figure}
The small tetrahedron (Fig.~\ref{smtetfig}) may be visualized as the result of a 
bisection of a medium tetrahedron.  
The form of the eigenvalue spectrum for the small tetrahedron is different from that for the 
previous two tetrahedra but the same as the cube's:
\be	
\lambda_{kmn}^2=\frac{\pi^2}{a^2}\left
(k^2+m^2+n^2\right)\,.
\ee
The Dirichlet case is the aforementioned ``pyramidal cavity" considered in Ref.~\refcite{AhmedovSmT}.
		
\subsubsection{Dirichlet BC}
The modal restriction for the complete set is $0<k<m<n$. The finite part obtained is thus
\begin{align}
E_{S}^{(D)}=\frac{1}{a}\bigg[& 
-\frac{1}{192\pi^2}Z_3(2;1,1,1)+\frac{1}{16\pi}\zeta(3/2)L_{-8}(3/2)+\frac{1}{32\pi}\zeta(3/2)\beta(3/2)\nn
\\&-\frac{\pi}{64}-\frac{\pi\sqrt{3}}{72}-\frac{\pi\sqrt{2}}{96}\,\bigg],
\end{align}				
which evaluates to
\be
E_{S}^{(D)} =-\frac{0.1005414622}{a}\,.
\ee	
This result differs from that of Ref.~\refcite{AhmedovSmT}. The discrepancy appears to stem from a 			
mode-counting error in Ref.~\refcite{AhmedovSmT}, and the result found there is likely wrong. 
		
\subsubsection{Neumann BC}
For the Neumann case, we again find the same condition that the mode numbers 
must satisfy $0\leq k \leq m \leq n$ excluding the origin. The Neumann 
Casimir energy is derived to be
\begin{align}
E_{S}^{(N)}=E_{S}^{(D)}-\frac{\zeta(3/2)}{16\pi a}\Big[2\,L_{-8}(3/2)
+\beta(3/2)\Big],
\end{align}		
with a numerical value of			 	
\be
E_{S}^{(N)} =-\frac{0.2587920021}{a}\,.
\ee	

\section{Analysis}
\begin{figure}[h]
\centering
\includegraphics[scale=0.7,  clip]{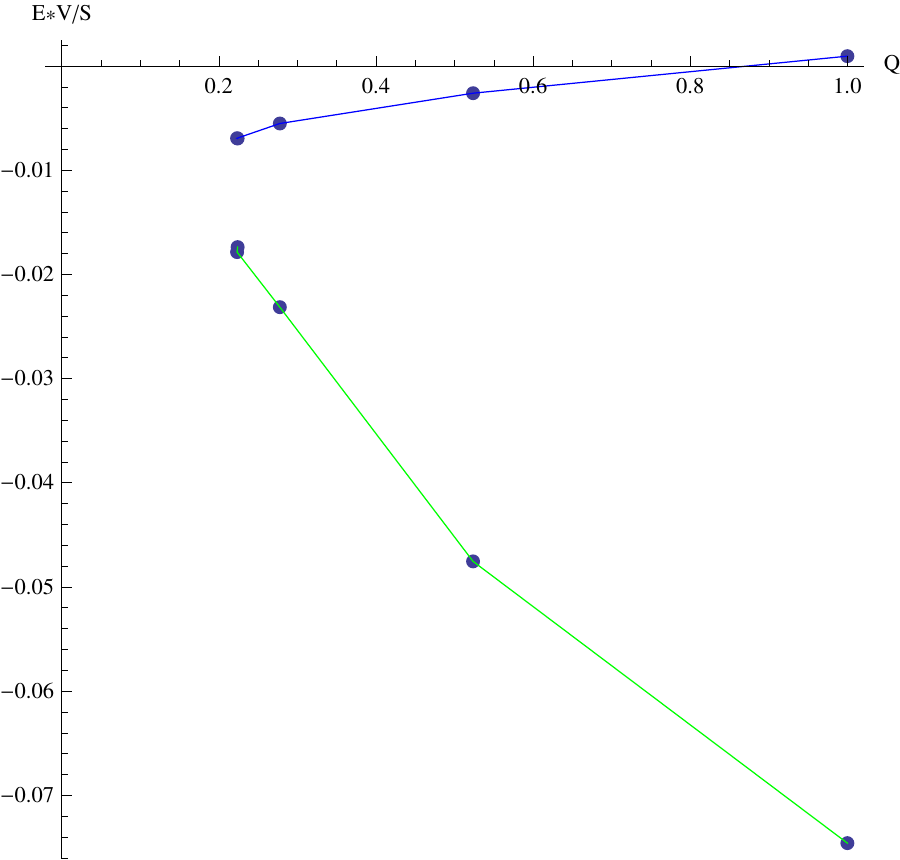}
\caption{Scaled Dirichlet and Neumann energies vs.\ isoareal quotients. The upper (lower) curve connects Dirichlet (Neumann) results. From left to right, 
the circular markers indicate small, medium (which cannot be resolved on this graph), 
large tetrahedra, cube, and sphere data points. It is interesting to note the relative order of the scaled energies of the small and medium tetrahedra: $E_M^{(D)}<E_S^{(D)}$ and $E_M^{(N)}>E_S^{(N)}$.}
\label{plot2dD}
\end{figure}

Seeking to gain a better understanding of self-energies, at least their correlation to the system's geometry, we add the well-known cases of cubic and spherical geometries.\cite{lukosz,DirichletSphere,Nesterenko} The scaled Casimir energies, 
$E_{\rm Sc}=E\times V/S$, are tabulated 
in Table \ref{tab1}, and
are plotted against the corresponding isoareal quotients, $\mathcal{Q}=36 \pi V^2/S^3$ in 
Fig.~\ref{plot2dD}. A noteworthy difference between the calculations of the energies for tetrahedra as compared
to a sphere is that in the polyhedral cases only the interior modes are considered 
(the exterior modes are unknown) whereas in the spherical cases both interior and exterior are 
(necessarily) included to cancel the curvature divergences.

\begin{table}[h]
\tbl{Scaled energies and isoareal quotients.}
{\begin{tabular}{@{}lccc@{}} \toprule
 & $\mathcal{Q}$ & $E_{\rm Sc}^{(D)}$ &
$E_{\rm Sc}^{(N)}$ \\
%& (Rad/s) & (Rad/s) \\ 
\colrule
Small T.      & 0.22327     &  $-0.00694$ & $-0.01787$ \\
Medium T.  & 0.22395 & $-0.00696$ & $-0.01739$\\
Large T.   & 0.27768  & $-0.00552$ & $-0.02315$\\
Cube   & 0.52359  & $-0.00261$ & $-0.04755$\\
Spherical Shell & 1 & \hphantom{$-$}0.00093 & $-0.07459$\\ \botrule
\end{tabular} \label{tab1}}
\end{table}

\section{Conclusions}
In this paper, we have extended the work of Ref.~\refcite{Cylpaper} from infinite cylinders to integrable tetrahedra.  The previous work was essentially two dimensional,
so it was possible to give closed form results for the Casimir self energy. 
This is  apparently not possible, at least not currently, for the cases considered in this paper.

The emerging systematics are intriguing, but not yet conclusive, since the cases we can evaluate
are limited.  Numerical work will have to be done to explore the geometrical dependence of
the Casimir energy of cavities composed of flat surfaces of arbitrary shape.

Work on other boundary conditions, in particular electromagnetic boundary conditions, is currently under way.
Unlike for cylinders, the electromagnetic energy of a tetrahedron is not merely the sum of Dirichlet
and Neumann parts; there is no break-up into TE and TM modes in general.  So this is a formidable task.
 Finite triangular prisms and polytopes are also currently the subject of ongoing work.
\section*{Acknowledgments}
We thank the US National Science Foundation and the US Department of Energy for
 partial support of this work.  We further thank Nima Pourtolami and Prachi
Parashar for collaborative assistance.

%\appendix

%\section{Appendices}

%\begin{thebibliography}{000} %for 3 digits

\end{document}